\documentclass[pra,twocolumn,showpacs]{revtex4-1}
\usepackage{graphicx}
\usepackage{bm}
\usepackage{amsmath}
\usepackage[percent]{overpic}
\usepackage{subfig}
\usepackage{amssymb}
\begin{document}

\title{There is no anomaly in the nonlocality of two entangled qutrits}
%\title{A measure of nonlocality which is maximal for maximally entangled qutrits}
\author{E. A. Fonseca} 
\author{Fernando Parisio}
\email{parisio@df.ufpe.br} 
\address{Departamento de F\'{\i}sica, Universidade Federal de Pernambuco, 50670-901,
Recife, Pernambuco, Brazil}

\begin{abstract}
There is no doubt about the fact that entanglement and nonlocality are distinct resources. It is acknowledged that a clear illustration of this point is the difference between maximally entangled states and states that maximally violate a Bell inequality. We give strong evidence that this anomaly may be an artefact of the measures that have been used to quantify nonlocality. By reasoning that the numeric value of a Bell function is akin to a witness rather than a quantifier, we define a measure of nonlocality and show that, for pairs of qutrits and of four-level systems, maximal entanglement does correspond to maximal nonlocality in the same scenario that gave rise to the discrepancy.
\end{abstract}
\maketitle
%
%\section{Introduction}
%
Entanglement \cite{plenio,horodecki} is behind some of the most perplexing physical effects ever observed. In spite of this, it alone, may be regarded as a purely mathematical concept: the failure of a vector in a Hilbert space ${\cal H}$ to be factorized as a single product of vectors in spaces ${\cal H}_i$, that together form ${\cal H}$ ($=\bigotimes_i{\cal H}_i$) \cite{comment}.  For mixed states there is a corresponding definition in terms of convex sums of explicitly separable density operators \cite{werner}.  Nonlocality \cite{mermin, brunnerB,popescuNP}, in contrast, also refers to the experimental scheme that is used to investigate the entangled state and, thus, to angles of Stern-Gerlach apparatuses or to the spatial disposition of beam splitters, for instance. 
That is to say, entanglement happens in Hilbert spaces while nonlocality manifests itself in our ordinary (3+1)D space (see however \cite{tipler}). Therefore, from a physical perspective, to carefully quantify nonlocality is as important as to seek entanglement measures. A faulty estimation of the extent of nonlocality embodied by a physical situation may lead to deceptive conclusions.

In an essay in honor of A. Shimony, Gisin provides a list of questions on Bell inequalities \cite{gisin1}. The one closely related to our goal in this work is ``Why are almost all known Bell inequalities for more than 2 outcomes maximally violated by states that are not maximally entangled?'' (for exceptions see \cite{jaksch, lim}). This fact originally reported in \cite{acin} and referred to as an anomaly \cite{methot} of nonlocality, has received a great deal of attention \cite{ji,jaksch,brunner,junge,hiesmayr,vidick}. To investigate this issue, we begin by addressing the tacit association often made between states that maximally violate a Bell inequality and maximally nonlocal states. 
 
Given a Bell function $I$, let us generically denote the associated inequality by
\begin{equation}
\label{bell}
I\le \xi\;,
\end{equation}
$\xi$ representing the bound imposed by local causality (we use this terminology instead of ``local realism'' \cite{gisinnote}). Then, if a state satisfies (\ref{bell}) for all possible settings of the measurement apparatus, it is local with respect to the inequality. Otherwise, the state is said to be nonlocal. Recall that the Werner matrices \cite{werner}, which are local with respect to the CHSH inequality, are nonlocal if more complex measurements are considered for dimension 5 or higher \cite{popescu} . A complete definition of locality must be exhaustive: a state is said to be local, with no further qualifiers, if it satisfies all relevant Bell inequalities. 

So, the notion of Bell nonlocality is quite neat.
The sensitive question is about {\it ordering}. What should one imply by asserting that state $\rho$ is more nonlocal than state $\sigma$?  
A common answer is that $\rho$ is more nonlocal than $\sigma$ if $I_{max}(\rho)$ is larger than $I_{max}(\sigma)$. The maxima being determined by scanning all possible settings. Although it is known that for any Bell function one can find another, equivalent function that arbitrarily increases the numerical value of the maximal violation \cite{brunnerB}, it is acknowledged that carefully normalised Bell inequalities may provide objective figures to quantify nonlocality. In what follows we reason against this view.

Insightful alternatives have been put forward in the last two decades. The tolerance of nonclassical correlations against noise has been considered as an operational measure of nonlocality \cite{dagomir1,ryu}, but this approach is not consensual \cite{acin}. In \cite{kl-div}, it is shown that optimal Bell tests occur for states that are neither maximally entangled nor maximally violating. The (statistically) optimal state found by the authors is the most suitable to unveil nonlocality, given that the experimentalist can only perform a finite number $N$ of realisations, as is always the case. However, at least in principle, one should be allowed to think of the limit $N\rightarrow \infty$, as we do with many other concepts in quantum theory. In this limit all nonlocal states can be safely devised. In a different framework, the communication cost for a local model to reproduce the quantum correlations has also been used as a task-based quantifier of nonlocality \cite{brassard, steiner, bacon}. However, different tasks usually induce different state orderings \cite{brunnerB}.

One can also ask what is the minimal detector efficiency required to evidence nonlocality for a given state. The interesting fact is that the efficiency required for the maximally entangled state of two qubits is larger than that of some partially entangled states \cite{eberhard, eberhard2} for the CHSH inequality \cite{chsh}. This might be considered an independent instance of the nonlocality anomaly. About this point, let us appeal to an analogy with entanglement. There is little doubt that the GHZ state is the maximally entangled state of 3 qubits, in particular, that it is more entangled than the $W$ state. However, in measuring the 3 particles on $|GHZ \rangle$, if the 3 detectors have an efficiency of $p < 1$, then the probability to witness the entangled character of $|GHZ \rangle$ is only $p^3$, since, whenever a single detector fails one only sees a maximally mixed, non-entangled ensemble for the remaining particles. In contrast, the probability to perceive entanglement in $|W \rangle$, with the same detectors, is $p^3+3(1-p)p^2\times 2/3$, since, by losing any of the particles we still have a noisy EPR pair (with a fidelity of $2/3$). Back to our problem, by the same token, although $(|00 \rangle+ |11\rangle)/\sqrt{2}$ is the most nonlocal, another state, only partially entangled, may present a nonlocality that is more resilient against detection inefficiency. {\it  The fact that a state is maximally nonlocal does not necessarily mean that its nonlocality is either the easiest  to detect or the most resistant against imperfections.} 

Recently, two measures have been experimentally implemented in \cite{experiment1}: the first one is based on how far is the state from the local polytope, and, the second is also related to the amount of communication needed to establish correlations. In \cite{vallone} a nonlocality quantifier has been defined, such that in certain scenarios it is inversely related to concurrence.
Some other proposals to quantify nonlocality can be found in the literature, focusing on multipartite systems \cite{branciard,bancal} and presenting nonlocality as a concept derivable from a notion of ``irreality''  \cite{renato}. 

Common to these previous works is the fact that the different figures of merit associated (or identified) to nonlocality attain their maxima for non-maximally entangled states. This does not violate any logical necessity, but, one should not refrain from a critical assessment of this, arguably, counterintuitive finding. In this work we define a measure of nonlocality which indicates that the anomaly that appears to exist for two entangled three- and four-level systems may well be an artefact of the previous definitions. Our suggestion seems to have deep, though simply definable, physical and statistical meanings.
%
%\section{Measure of Bell Nonlocality}
%

A tenable reasoning about the quantification of nonlocality is that some clue might come from nonlocal hidden variable (NLHV) models capable of reproducing the quantum correlations. For example, one could say that a state is more nonlocal than another if the underlying NLHV model violates local causality in different degrees for these different states. This, however, cannot be inferred from these models in any obvious way. The distance between subsystems does not enter in the Bell functions, after all. As a first illustration we refer to the model developed by Bell in his seminal paper \cite{bell}. It is simply assumed that ``the results of measurements with one magnet now depend on the setting of the distant magnet [...]''. The mutual influence between the subsystems being instantaneous, no matter the numerical value assumed by the Bell function. In a more general picture, consider the NLHV theory {\it par excellence},  Bohmian mechanics \cite{bohm}. The so-called quantum potential does not react faster or slower, for different states, under a measurement on one of the subsystems. In particular, this holds for two entangled rotors of spin-1/2 \cite{ramsak}. The fact that, for a given Bell inequality, some of these instantaneous interactions are related to non violating states must be understood in the light of the generalised Gisin's theorem: all bipartite $N\times N$ entangled states violate some Bell inequality \cite{teogisin,chen}.
The action at a distance appears to be equally spooky for all nonlocal states within these NLHV models. 

Even for theories relying on finite (superluminal) signalling speed, the relation $I_{max}(\rho)>I_{max}(\sigma)>\xi$ does not necessarily imply that $v_{\rho}>v_{\sigma}>c$, where $v$ is the signal velocity for each state and $c$ is the speed of light. This reasoning suggests that all violating states {\it for a particular setting} are equally nonlocal, that the essential information provided by a Bell inequality is of a seemingly Boolean nature, a state being either local or nonlocal with respect to those settings, without gradations.

This apparently all-or-nothing picture, however, does not lead to a dead-end. On the contrary, it points to a conceptually simple solution. 

Given a state and a specific Bell inequality, the most exhaustive experiment one can go through is to investigate local causality for all settings. For simplicity of presentation we refer to inequalities associated to non-degenerate von Neumann measurements \cite{comment2}. Based on our previous discussion, we are lead to state that $\rho$ is more nonlocal than $\sigma$ if the former violates the inequality, {\it by any extent}, for a larger amount of setting parameters than the latter. This statement can be cast in very simple statistical terms:  $\rho$ is more nonlocal than $\sigma$ if, for an unbiased random choice of settings, the probability to obtain a violation is larger for $\rho$.

To formalise this idea, we define the space ${\cal X}=\{x_1, \dots ,x_n \}$ of all possible parameters determining the settings for a given (preferably tight) Bell inequality. For a particular state $\rho$, let ${ \Gamma}_{\rho} \subset {\cal X}$ be the set of points that lead to violation and $V({\rho)}$ be proportional to the volume of ${ \Gamma}_{\rho}$. We say that if $V(\rho)>V(\sigma)$, then $\rho$ is more nonlocal than $\sigma$, with
\begin{equation}
\label{measure}
V(\rho)\equiv \frac{1}{\cal N}\int_{{ \Gamma}_{\rho}}{\rm d}^n x\;,
\end{equation}
where ${\cal N}$ is a normalisation constant. The measure of integration is such that every setting (set of parameters) has equal weight. For instance, one setting corresponding to a direction in space demands two parameters one polar ($\varphi$) and one azimutal ($\theta$) angle, leading to ${\rm d}^2 x={\rm d}\Omega=\sin \theta {\rm d}\theta{\rm d}\varphi$. If, on the other hand, the settings are defined by the plane angles of $n$ polarisers, e.g., then we simply have ${\rm d}^nx={\rm d}\varphi_1\dots {\rm d}\varphi_n$. We call $V$ the {\it volume of violation}.
Hereafter we focus on the important case where the settings are such that ${\cal X}$ is a bounded set. We remark that the numeric calculations needed to determine the volume of violation are the paradigmatic problem for which Monte Carlo methods are intended \cite{montecarlo}. The above definition has no relation to the volume of the set of separable states defined in \cite{karol}, the volume of violation is an integration over the experimental parameters that can be varied within the context of a given Bell inequality.

A more fundamental definition should not invoke a particular Bell inequality, but rather, the set of conditional probabilities $P(ab | xy)$ (also called behaviors), where $a$ and $b$ are outputs and $x$ and $y$ are inputs, see \cite{vicente}. This amounts to an integration similar to (\ref{measure}), but over the exterior, no signalling part of the local polytope, which, however is an exponentially hard computational problem. In addition, in this first account we intend to address the very same situation that gave rise to the anomaly. In \cite{vicente}, some criteria are given that reasonable measures should fulfil in terms of operations in the space of behaviours. In this more general picture one interesting question is whether $V$ satisfies those criteria.

\begin{figure}[ht!]
\includegraphics[width=4.27cm,angle=0]{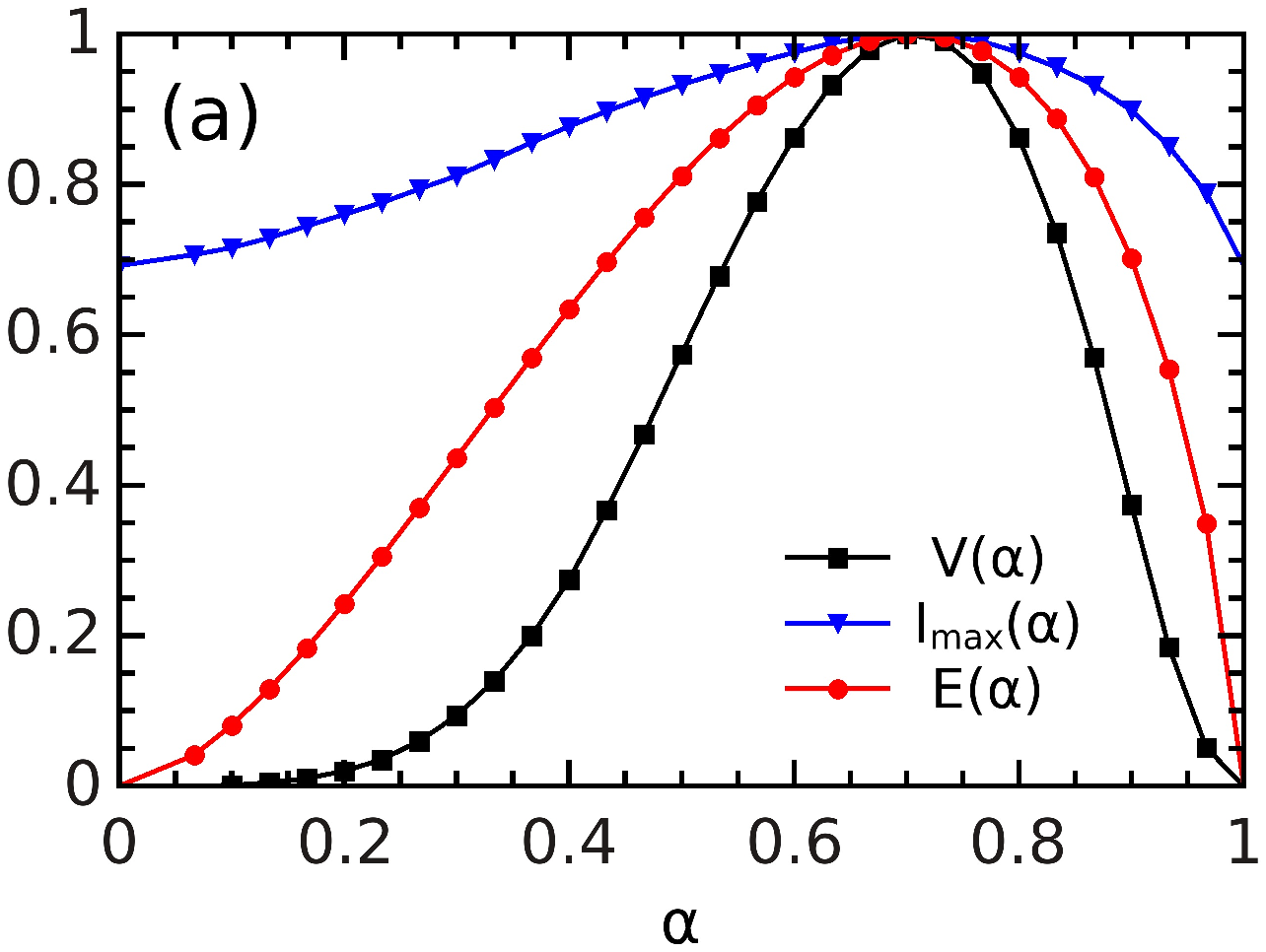}
\includegraphics[width=4.27cm,angle=0]{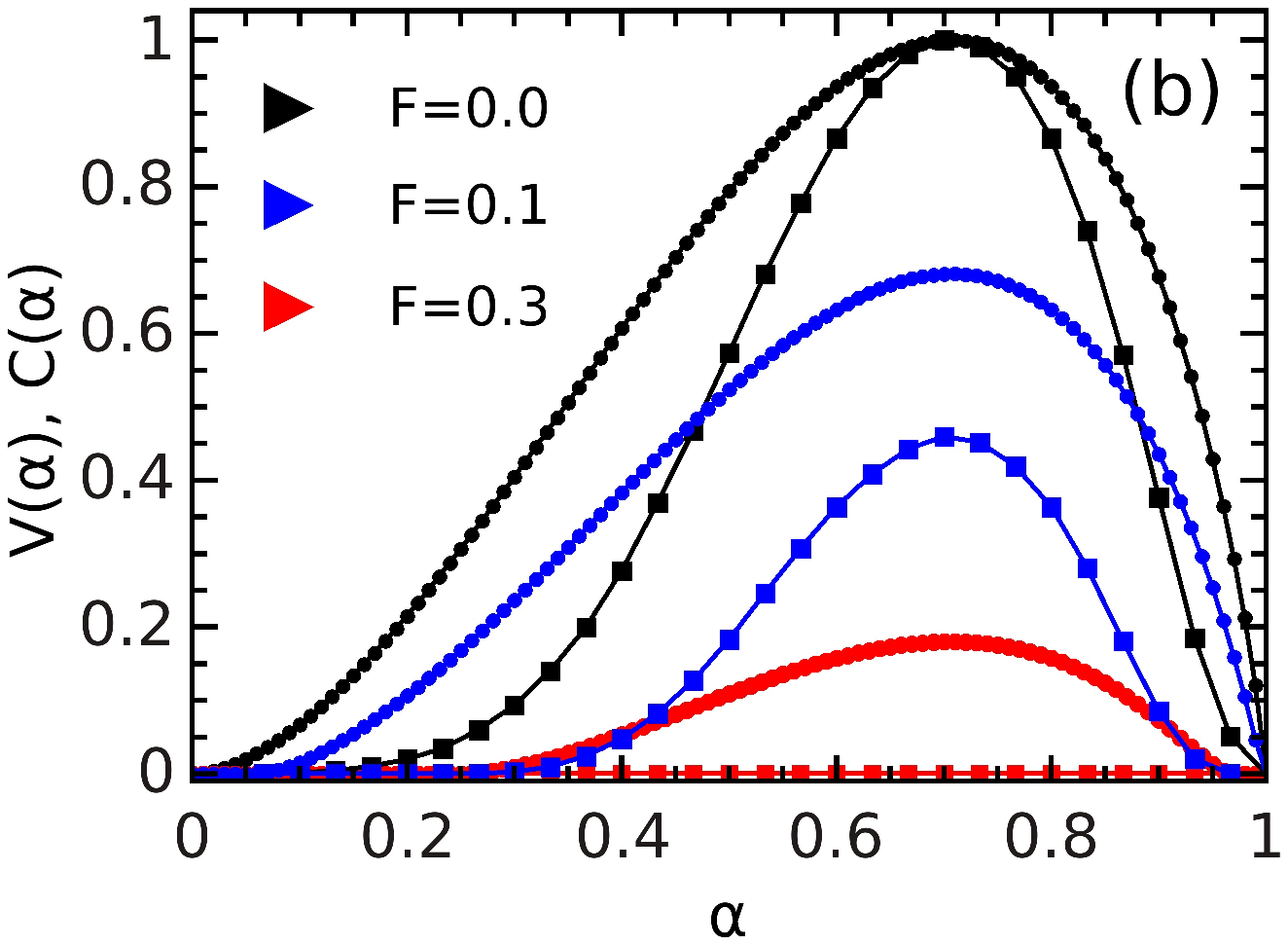}
\caption{(color online) In (a) we show the entropy of entanglement (circles), ${I}_{max}$ (triangles), and $V$ (squares) versus $\alpha$ for the pure state $|\psi_{\alpha} \rangle$. In (b) noise is considered and the plots correspond to $V(\alpha)$ compared to the concurrence $C(\alpha)$ for different values of the noise fraction $F$. }
\label{figure1}
\end{figure}

As an initial test, we consider the CHSH inequality \cite{chsh} for two entangled qubits in pure and mixed states. In this case the Bell function depends on four unit vectors: $I_{CHSH}(\hat{a},\hat{b}, \hat{c}, \hat{d})= |E(\hat{a},\hat{b})-E(\hat{a},\hat{d})|+E(\hat{c},\hat{d})+E(\hat{c},\hat{b})$, with $E$ being a correlation function defined for a pair of directions.
We can write it more explicitly in terms of eight angular parameters, $I_{CHSH}=I(\theta_a, \varphi_a,\theta_b, \varphi_b, \theta_c, \varphi_c,\theta_d, \varphi_d,)$, ${\cal X}$ corresponding to the cartesian product of four unit spheres, yielding ${\rm d}^n x = {\rm d}\Omega_a{\rm d}\Omega_b{\rm d}\Omega_c{\rm d}\Omega_d$, with ${\rm d}\Omega_i=\sin \theta_i {\rm d}\theta_i {\rm d}\varphi_i$. We found that the maximally entangled state maximises, both, $I$ and $V$. In Fig. \ref{figure1} (a) we show these quantities along with the entropy of entanglement for the family of pure states 
\begin{equation}
|\psi_{\alpha}\rangle=\alpha |00\rangle+\sqrt{1-\alpha^2}|11\rangle\;,
\end{equation}
as functions of $\alpha$. The volume $V$ is rather sensitive to variations of $\alpha$, presenting the steepest descent from its maximum at $\alpha = 1/\sqrt{2}$. In Fig. \ref{figure1} (b) we plot the concurrence $C(\alpha)$ \cite{hill,wooters} and $V(\alpha)\equiv V(\rho_{\alpha})$ of the noisy state $\rho_{\alpha}=(1-F) |\psi_{\alpha} \rangle \langle\psi_{\alpha} |+F\;\mathbb{I}/4 $, where $\mathbb{I}$ is the $4\times 4$ identity operator and $F$ is the noise fraction. The volume of violation is more fragile against noise than entanglement. Around a noise fraction of $F\approx0.3$, nonlocality, as rendered by $V$, completely disappears.
We also applied our measure to the first Bell \cite{bell} inequality (CHSH with $\hat{c}=\hat{d}$) and, importantly, to the inequality 3322 \cite{3322} (inequivalent to CHSH), with similar results.
So far, the volume of violation gives no sensible new information in comparison to the maximum of the Bell functions, yet, it is consistent with our expectations on what should be a nonlocality measure in the safe terrain of two entangled qubits.
This agreement between $V$ and $I_{max}$ ceases to happen when two higher dimensional systems are considered, even in the pure case.
%
%\section{Two qutrits and beyond}
%

Now we consider two entangled qutrits, i. e., a composite system with Hilbert space ${\cal H}={\cal H}_1\otimes {\cal H}_2$, dim${\cal H}_1$=dim${\cal H}_2$=3. Let $\{|0\rangle_i, |1 \rangle_i, |2\rangle_i\}$ be an orthonormal basis in ${\cal H}_i$ ($i=1,2$) and consider the three-outcome observables $A_a$, for system 1, and $B_b$ for system 2 ($a,b=1, 2$). It has been shown in \cite{collins} that local hidden variable models must satisfy the tight inequality
\begin{eqnarray}
\label{I3}
\nonumber
I_3=P(A_1=B_1)+P(B_1=A_2+1)+P(A_2=B_2)\\
\nonumber 
+P(B_2=A_1)-P(A_1=B_1-1)-P(B_1=A_2)\\
-P(B_2=A_1-1)\le 2\;,
\end{eqnarray}
where the arguments of the probabilities above are taken modulo 3, e.g., $P(B_1=A_2+1)= P(B_1=0,A_2=1)+P(B_1=1,A_2=2)+P(B_1=2,A_2=0)$. As in \cite{acin} let us focus on the family of pure states 
\begin{equation}
\label{state}
| \Psi_{\gamma} \rangle=\frac{1}{\sqrt{2+\gamma^2}}(|00\rangle+\gamma |11\rangle+|22\rangle)\;.
\end{equation}
The maximal entropy of entanglement is, naturally, given by $\gamma=1$, while it was shown that (\ref{I3}) is maximally violated by the state with $\gamma \equiv \tilde{\gamma}\approx 0.792$  \cite{acin}. Entanglement and Bell nonlocality are, indeed, physically distinct, but, the fact is that the former constitutes the sole source of the latter (we exclusively refer to Bell nonlocality \cite{bennett}). Thus, it is not unreasonable to expect that the maxima should coincide.

Let us apply measure (\ref{measure}) to this problem. Note carefully that general Stern-Gerlach-type measurements on a pair of spin-1 particles only demand eight parameters. However, these measurements do not reveal the whole richness of the Hilbert space \cite{dagomir1}.  For this reason, in order to calculate $V(\gamma)\equiv V(| \Psi_{\gamma} \rangle\langle  \Psi_{\gamma}|)$ one must perform an integration in a twelve-dimensional space, as we will see.
General unitary operations are achievable in the laboratory via multiport beam splitters \cite{klyshko,reck,zukowski}. In this optical context the whole space of parameters can be visited by varying the reflectivity of beam splitters and the angle of phase shifters, for instance. 
\begin{figure}[!ht]
    \subfloat{%
      \begin{overpic}[width=0.52\textwidth]{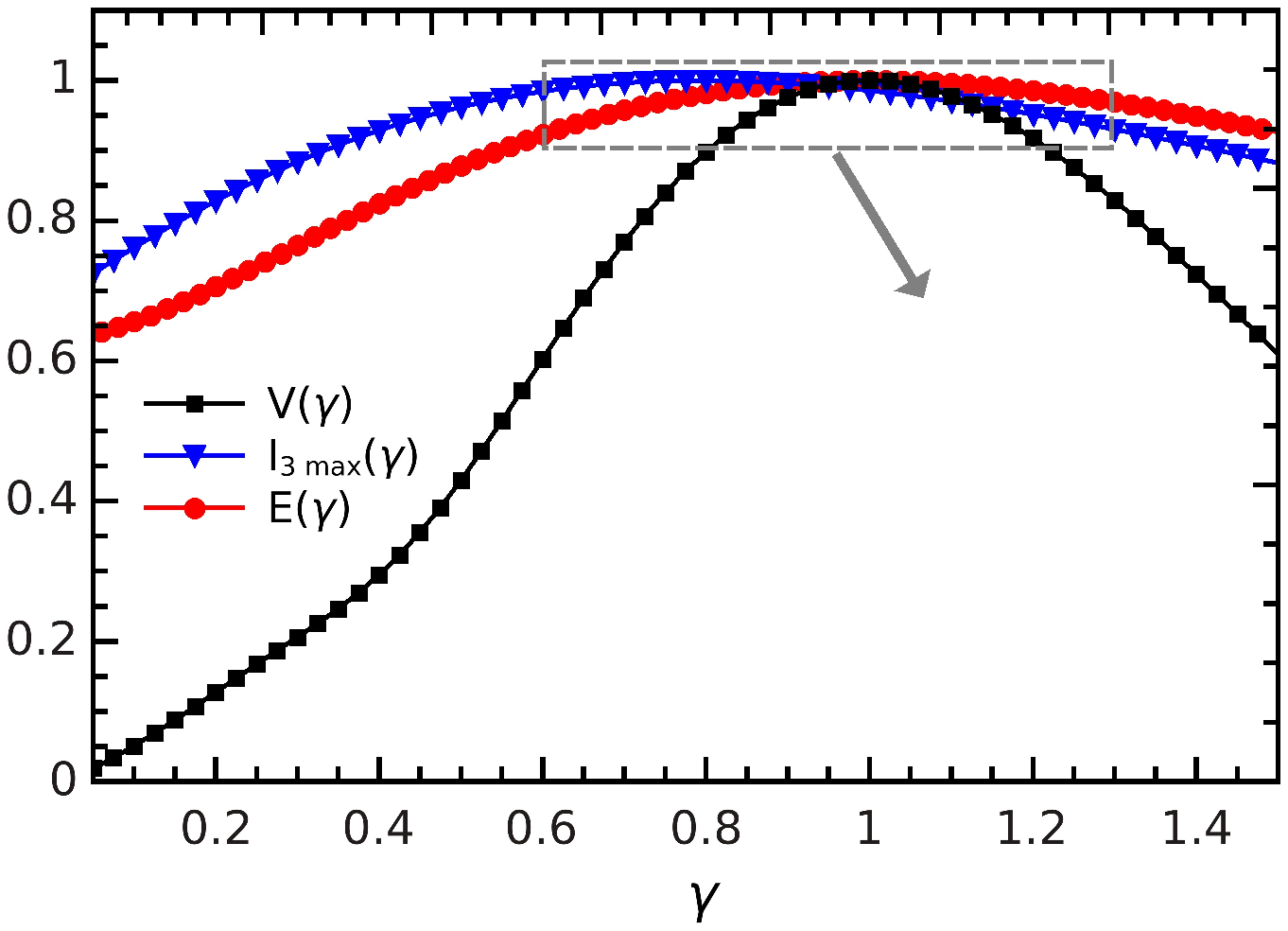}
        \put(43,12){\includegraphics[width=46\unitlength]{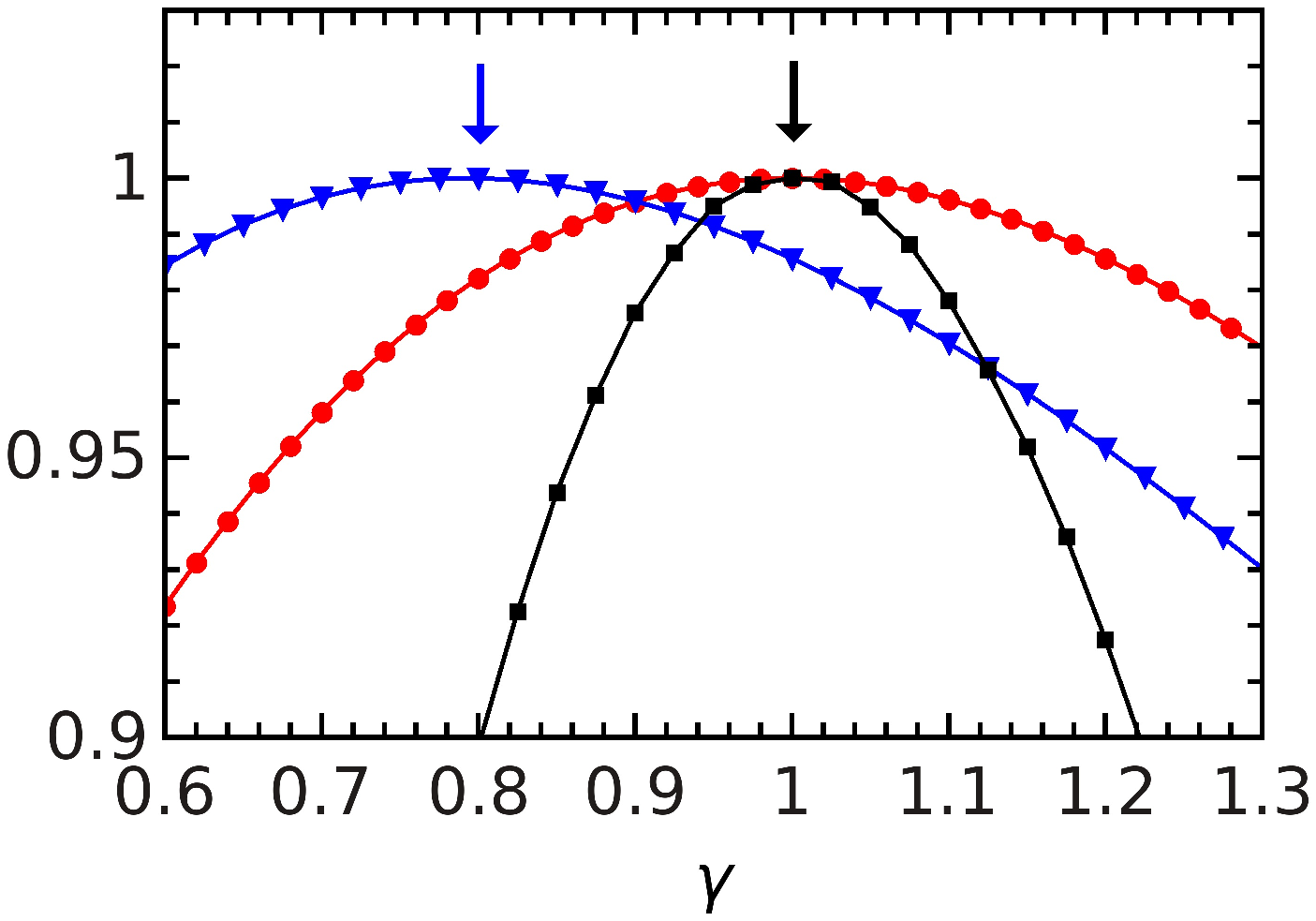}}
      \end{overpic}
    }
    \caption{(color online)  Entropy of entanglement (circles), ${I_3}_{max}$ (triangles), and $V$ (squares) as functions of $\gamma$ for state (\ref{state}). All quantities are normalised such that their maximal value is 1. The inset shows a zoom in of the region marked by the rectangle in dashed lines.}
\label{figure2}
\end{figure}
From the family of states (\ref{state}), with these linear optical elements, we can get:
\begin{equation}
\label{stateb}
| \Psi' \rangle=\frac{1}{3}\sum_{j,k,l=0}^{2}\alpha_j e^{i[\phi_a(j)+\varphi_b(j)]} e^{i\frac{2\pi}{3} j(k+l)}|k\;l\rangle\;,
\end{equation}
with $a,b=1, 2$, and $\alpha_0=\alpha_2=1$, $\alpha_1=\gamma$. The optimal parameters for violations of (\ref{I3}) by the maximally entangled state have been determined in \cite{collins,dagomir} and reads $\phi_1(j)=0$, $\phi_2(j)=\pi j/3$, $\varphi_1(j)=\pi j/6=-\varphi_{2}(j)$. The puzzling situation arises when one sees that the maximal violation has a peak at $\gamma = \tilde{\gamma}$.

In Fig. \ref{figure2} we compare our numeric calculations of $V(\gamma)$ to the normalised entropy of entanglement $E$ and to the maximum of $I_3$. The maxima of $E$ and $V$ coincide exactly at $\gamma=1$, as can be seen in the inset, while ${I_3}_{max}$ attains its maximum at $\gamma=\tilde{\gamma}$. This shows that the anomaly in the nonlocality of two entangled qutrits does not exist, if one adopts the volume of violation as the measure of nonlocality.

\begin{figure}[ht!]
\includegraphics[width=4.25cm,angle=-0]{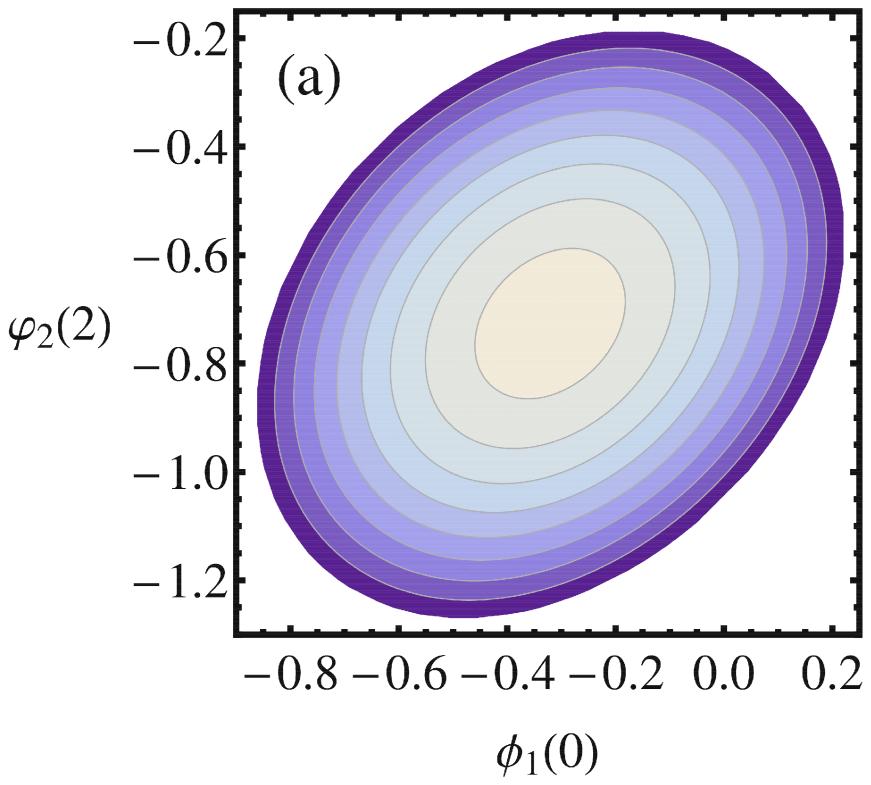}
\includegraphics[width=4.25cm,angle=-0]{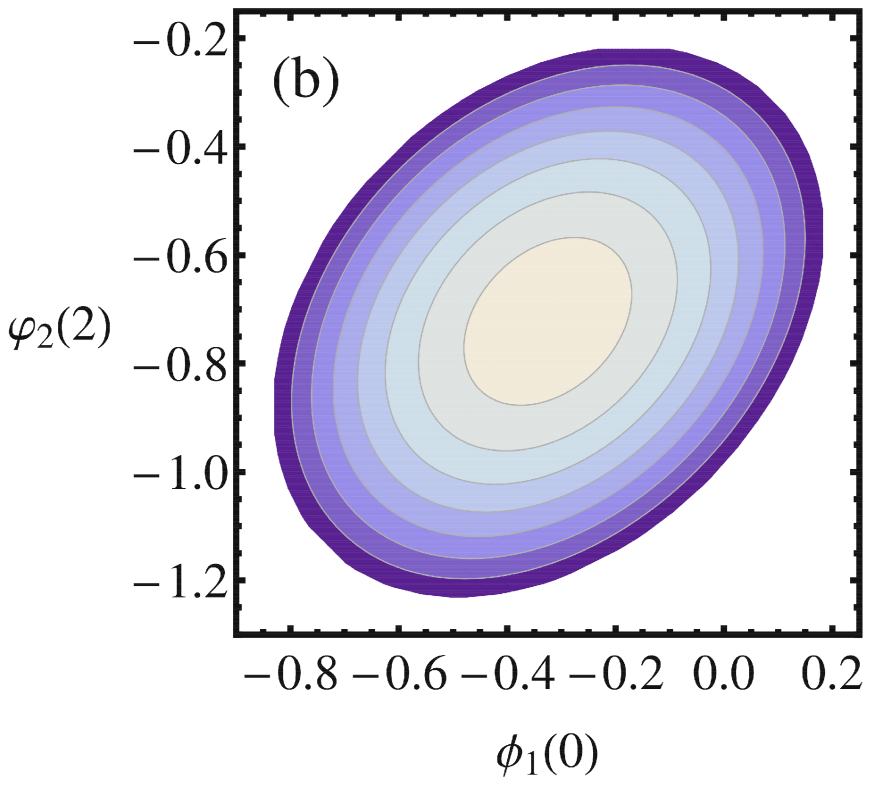}
\caption{(color online) Sections $\phi_{1}(0)-\varphi_{2}(2)$ of the 12-dimensional space ${\cal X}$. Some parameters were set away from the optimal values. The area of violation for $\gamma=1$ (a) is $14\%$ larger than that for $\gamma=0.792$ (b).}
\label{figure3}
\end{figure}

It is easy to understand what is going on. Although 
$|\Psi_{\tilde{\gamma}}\rangle$ presents a more pronounced maximum of ${I_3}_{max}$ in comparison to $|\Psi_{1}\rangle$, the nonlocality of the former is less robust, for, as we get farther away from the optimal setting in ${\cal X}$, $I_3(\Psi_{\tilde{\gamma}})$ falls off faster than $I_3(\Psi_{1})$. This effect on the volume of violation is clearly illustrated in Fig. \ref{figure3}, where two-dimensional sections [$\phi_1(0)-\varphi_2(2)$] of $\Gamma$ are shown for $|\Psi_{1}\rangle$ (a) and for $|\Psi_{\tilde{\gamma}}\rangle$ (b). The other parameters are set as $\phi_2(0)=\phi_2(1)=\pi j/6$, $\varphi_1(j)=0$, the remaining angles keeping the optimal values. In this particular example the violation area for $\gamma=1$ is about 14 $\%$ larger than that for $\gamma=\tilde{\gamma}$. The scales are identical in both figures. 
Finally, to be sure that this conciliation between the maxima of entanglement and nonlocality is not an unlikely coincidence, we addressed the problem of two four-dimensional Hilbert spaces. We considered the following family of entangled states
\begin{equation}
\label{stateN4}
| \Psi_{\lambda_1,\lambda_2} \rangle=\frac{1}{\Lambda}(|00\rangle+\lambda_1 |11\rangle+\lambda_2|22\rangle+|33\rangle)\;,
\end{equation}
with $\Lambda=\sqrt{2+\lambda_1^2+\lambda_2^2}$. The CGLMP inequality is maximally violated by a state that is not maximally entangled, given by $\lambda_1=\lambda_2\approx 0.739$, yielding $I_4\approx 2.973$. We surveyed the volume of violation associated to $I_4$ in the region $(\lambda_1\, ,\lambda_2) \in [0.6, 1.2] \times [0.6,1.2]$. Once again, $V$ is maximal for $\lambda_1=\lambda_2=1$ among all investigated states. In particular, the ratio of the volumes $V$ of the maximally entangled and maximally violating states is around $1.24$. Exhaustive numerical investigations in in the whole Hilbert spaces in the spirit of \cite{laskowsky} would be very welcome. However, these demand a large amount of computational resources.
%
%\section{Conclusion and perspectives}
%

We argue that, given a state, a Bell inequality, and a {\it particular} setting, there should be no gradations of nonlocality, the inequality functioning as a witness. However, by ``tracing over the settings'', attributing equal weight to all those that violate the inequality and weight zero to those that do not lead to violations, we showed that it is possible to quantify Bell nonlocality in a consistent way. In particular, within the context of our proposal, there is no discrepancy between maximally entangled and maximally nonlocal states, at least for entangled qutrits and also for systems composed of two four-level subsystems.

In this work the normalisation constant ${\cal N}$ in Eq. (\ref{measure}) played no important role. We simply set it such that $0\le V \le 1$, with $V=1$ for the maximally nonlocal pure state. We remark that there is a more absolute definition, which, however, would make the presented results a little cumbersome to analyse. This definition is ${\cal N}=($vol. of ${\cal X})$, leading to $V(\rho)=($vol. of $\Gamma_{\rho})/($vol. of ${\cal X})$. In this way the volumes of violation associated to the same state but different inequalities can be numerically compared.

An interesting point, that is presently under consideration, is the possibility to compare the nonlocality of the maximally entangled state as the dimension of the Hilbert space varies. In \cite{dagomir1} it is stated that violations in the principle of locality are stronger for two qunits than for two qubits. This conclusion was based on resistance to noise and, thus, an interesting question is wether or not we can reach a similar conclusion by employing the volume of violation.

Another question that has gained relevance is why quantum mechanics presents weaker nonlocality than ``probability boxes'', like PR-boxes \cite{pr}? Under the light of our results, the statement implied in the question should be reassessed, since it is based solely on the numerical value associated to the violation for a particular setting. It is an exciting perspective to check whether or not the volume of violation of the ``super singlet'' presented in \cite{pr}, in fact supports a stronger nonlocality.

To obtain the results described after Eq. (\ref{stateN4}) an integration (split in a cluster with 20 cores) in a 16-D space took a couple of days. We, thus, finish with the hope that analytical properties of $V$, Eq. (\ref{measure}), can be derived, helping to reduce the computational effort to calculate the volume of violation involving states in Hilbert spaces of higher dimensions.
\begin{acknowledgements}
We thank R. M. Angelo for his suggestions on several aspects of this work. Financial support from Conselho Nacional de Desenvolvimento Cient\'{\i}fico e Tecnol\'ogico (CNPq) via the Instituto Nacional de Ci\^encia e Tecnologia - Informa\c{c}\~ao Qu\^antica (INCT-IQ), Coordena\c{c}\~ao de Aperfei\c{c}oamento de Pessoal de N\'{\i}vel Superior (CAPES), and Funda\c{c}\~ao de Amparo \`a Ci\^encia e Tecnologia do Estado de Pernambuco (FACEPE) is acknowledged.
\end{acknowledgements}


\begin{thebibliography}{99}
\bibitem{comment} By adopting this view one should consistently agree that task-based entanglement measures, 
although insightful, quantify some other resource, closely related to entanglement, but not identical to it.
\end{thebibliography}
\end{document}